\documentclass[twocolumn,epsf,prb,amsmath,amssymb,showpacs]{revtex4}
\usepackage{graphicx}

\begin{document}
\title{Landau levels and oscillator strength in a biased bilayer of graphene}
\author{J. Milton Pereira$^{1,2}$ Jr., F. M. Peeters$^1$, and P. Vasilopoulos$^3$}
\address{$^1$Department of Physics, Universiteit Antwerpen Groenenborgerlaan 171, B-2020 Antwerpen, Belgium\\
$^2$Departamento de F\'{\i}sica, Universidade
Federal do Cear\'a, Fortaleza, Cear\'a, $60455$-$760$, Brazil\\
$^3$Department of Physics, Concordia University, Montreal, Quebec,
Canada H3G 1M8}

\begin{abstract}
We obtain analytical expressions for the eigenstates and the
Landau level spectrum of biased graphene bilayers in a magnetic
field. The calculations are performed in the context of a
four-band continuum model and generalize previous approximate
results. Solutions are presented for the spectrum as a function of
interlayer coupling, the potential difference between the layers
and the magnetic field. The explicit expressions allow us to
calculate the oscillator strength and the selection rules for
electric dipole transitions between the Landau states.
Some transitions are significantly shifted in energy relative to
those in an unbiased bialyer and exhibit a very different magnetic
field dependence.
\end{abstract}
 \maketitle

\section{Introduction}

The recent realization of stable single layer and bilayer carbon
crystals (graphene) has aroused considerable interest in the study
of their electronic properties
\cite{zheng,novo3,novo4,shara,zhang}. These materials have been
shown to display interesting new phenomena that may allow the
development of carbon-based nanoelectronic devices. The behavior
of charge carriers in wide single-layer graphene sheets has been
described as that of ultrarelativistic massless particles, with a
"light speed" equal to the Fermi velocity of the crystal and a
gapless linear dispersion close to the $K$ and $K'$ points. One
consequence of that is 
that single-layer graphene
displays an unusual quantum Hall effect, in which the quantum Hall
plateaus are found in half-integer multiples of $4\,e^2/h$.
Moreover, the massless character of the charge carriers in
single-layer graphene gives a $\sqrt{B}$ dependence to the Landau
levels (LL), which has been recently confirmed by infrared
transmission and cyclotron resonance experiments in thin graphite
samples \cite{Potemski} and on single layers of graphene
\cite{Deacon}. On the other
hand, for an unbiased graphene bilayer  the spectrum at the vicinity of the $K$ points displays four
parabolic bands but is still gapless \cite{Bart,Ohta1}. The
absence of a gap, together with the chiral nature of the
electronic states, in both single-layer and bilayer graphene, is at
the root of phenomena such as the Klein tunneling
\cite{Katsnelson,Milton2,Che} which has important consequences for
the design of graphene-based devices \cite{Milton3}.

Recent theoretical and experimental results have shown that a gap
can be induced in bilayer graphene by changing the density of
charge carriers in the layers through the application of an
external field or by chemical doping, which creates a potential
difference between the layers \cite{Ohta,Min}. The presence of
such a bias transforms the graphene bilayer into a semiconductor
with a tunable gap. This implies that such biased bilayers can be
used to confine charge carriers through the use of, e.g.,
nanostructured gates or spatially varying doping \cite{Milton}.
Tight-binding calculations have shown that the existence of such a
gap can have a significant effect on the LL spectrum of biased
graphene bilayers \cite{McCann,Castro}

In this work we determine the spectrum of a biased graphene
bilayer in the presence of an external magnetic field in the
continuum description and obtain, by solving the four-band
Hamiltonian, the exact eigenstates as well as the oscillator
strengths and the selection rules for the electric dipole
transitions between LL. Previous numerical results for the LL in
biased and unbiased bilayer graphene \cite{Falko,Castro}, were
obtained within a model based on a reduced two-band Hamiltonian,
which is accurate for energies much lower than the interlayer
coupling parameter. Furthermore, these previous studies used a
description in which the Hamiltonian can be described
approximately in terms of ladder operators in the basis of Landau
functions, which is only possible in the Landau gauge. Here, we
generalize these results to much higher energy by using the more
accurate four-band model, and obtain exact expressions in both the
Landau gauge and the symmetric gauge. The calculations allow us to
obtain a significant correction to the spectrum obtained by
previous approximate methods. The model is presented in Sec. II
for zero magnetic field $B$. A finite field $B$ is considered in
Sec. III, in the Landau and symmetric gauges, and the results are
given in Sec. IV. The oscillator strength is evaluated in Sec. V
and a summary follows in Sec. VI.

\section{Zero Magnetic Field} The crystal structure of an undoped bilayer of
graphene is that of two honeycomb sheets of covalent-bond carbon
atoms coupled by weak Van der Waals forces. To each carbon atom
corresponds a valence electron and the structure can be described
in terms of four sublattices, labelled A, B (upper layer) and
A$^{\prime}$, B$^{\prime}$ (lower layer).
In this work we focus on the Bernal stacking, in which half the
atoms of the upper layer (i.e. the sites of the A sublattice) are
on top of half the sites of the lower layer (the sites of the B'
sublattice). This type of stacking is found in graphite and has
been reported as well for samples of bilayers of graphene.
Alternatively, one can consider the hexagonal stacking, in which
all sites of the top layer are on top of all sites of the lower
layer. That configuration introduces a different coupling scheme
between the layers, and can lead to qualitatively distinct
results. However, the band structure data from the epitaxial
bilayer samples of ref. 15, as well as the results from the
samples produced by micro-mechanical cleavage of graphite in ref.
21 are consistent with the Bernal stacking.
The coupling between the layers is described by an interaction
term
between
the $A$ and $B'$
sublattice sites. Considering only nearest-neighbor hopping, the
Hamiltonian of the system in the vicinity of the $K$ point is
given, in the continuum approximation, by \cite{Snyman}
\begin{equation}
\mathcal{H}=\mathcal{H}_0 +(\Delta U/2)\tau_z,
\end{equation}
with
\begin{equation}
\mathcal {H}_0=
\begin{pmatrix}
  U_0 & \pi & t & 0 \\
 \pi^\dagger & U_0 & 0 & 0\\
  t & 0 & U_0 & \pi^\dagger\\
 0 & 0 & \pi & U_0
\end{pmatrix}
;
\end{equation}
the operator $\tau_z$ assigns a positive (negative) sign to
the upper (lower) layer labels and is defined as
\begin{equation}
\tau_z=
\begin{pmatrix}
  \mathbf{I} & 0  \\
 0 & -\mathbf{I}
\end{pmatrix}
,
\end{equation}
with $\mathbf{I}$ denoting the $2\times 2$ identity matrix. We
assume a
constant
 interlayer coupling term $t \approx 400$ meV; $\pi =
v_F(p_x + ip_y)$, $\hat{\mathbf p} = (p_x,p_y)$ is the 2D momentum
operator, $v_F \approx 1\times 10^6$ m/s, $U_0 = (U_1+U_2)/2$,
$\Delta U = U_1 - U_2$ and $U_{1}$, $U_{2}$ are the potentials at
the two layers, which reflect the influence of doping on one of
them and/or the interaction with an external electric field, i.e.,
the gating. In this work we treat the potentials at each layer and
the gap as adjustable parameters, which may be obtained
experimentally, e.g. from angle-resolved photoemission
spectroscopy measurementes \cite{Ohta}. The eigenstates of Eq. (1)
are four-component spinors $\Psi = [\psi_A \, , \, \psi_B\, , \,
\psi_{B'}\, , \, \psi_{A'}]^T$, where $\psi_{A,B}$
($\psi_{A',B'}$) are the envelope functions associated with the
probability amplitudes at the respective sublattice sites of the
upper (lower) graphene sheet. The superscript $T$ denotes the
transpose of the $[...]$ vector.

If the magnetic field is absent and the potentials $U_1$ and $U_2$
are constant, the single-particle spectrum consists of four bands
with eigenvalues \cite{Milton}
\begin{eqnarray}
&&E^+_{\pm}(k) = U_0+(1/2)\big[( t \pm
\Gamma)^2+\Omega\big]^{1/2},\\ &&\cr &&E^-_{\pm}(k) =
U_0-(1/2)\big[( t \pm \Gamma)^2+\Omega \big]^{1/2},
\end{eqnarray}
where $ \Gamma =\big[t^2+4s_{F}^2 + 4(s_{F}^{2}/t^2)\Delta
U^2\big]^{1/2}$, $s_{F}=\hbar v_F k$, and $\Omega =
\big[1-4s_{F}^2/t^2\big ]\Delta U^2$. Note that  for $k=0$ the
spectrum shows a gap at $k=0$ of size $E^+_-(0)-E^-_-(0)=|\Delta
U|$ and the system becomes a narrow-gap semiconductor. Figure 1
shows the low energy electronic spectrum of graphene bilayers for
$U_0 = 0$ and three different values of gap: $\Delta U = 0$ (black
solid line), $\Delta U = 25$ meV (red dashed line), $\Delta U =
100$ meV (blue dotted) and $\Delta U = 200$ meV (green dot-dashed
line). For large $k$-values the linear $E-k$ behavior is
recovered. The hole spectrum in this case is obtained by taking
$E(k) \rightarrow -E(k)$. In the absence of bias, the spectrum
shows approximately parabolic bands, which become increasingly
deformed as $\Delta U$ increases, with the appearance of energy
minima at non-zero values of $k$, and can be approximated by a
fourth-order polynomial function \cite{Bart,Stauber}. As we show
below, this non-parabolic dispersion has a significant influence
on the LL results.
\begin{figure}
\vspace*{-0.95cm}
\includegraphics*[height=8cm, width=8.5cm]{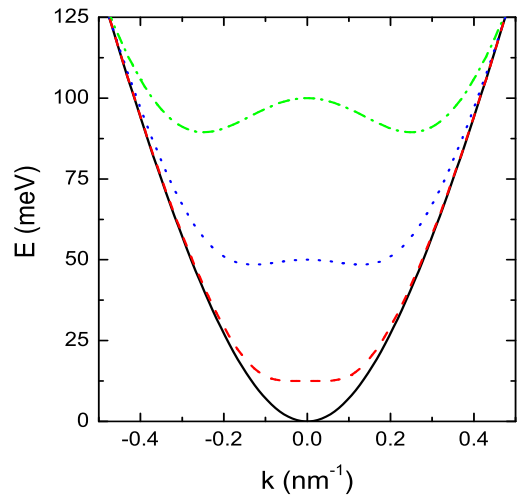}
\vspace*{-0.8cm} \caption{Low-energy electronic dispersion of
graphene bilayers for $B=0$ for different values of the gap,
namely $\Delta U = 0$ (black solid line), $\Delta U = 25$ meV (red
dashed line), $\Delta U = 100$ meV (blue dotted) and $\Delta U =
200$ meV (green dot-dashed line). In all cases $U_0 = 0$.}
\end{figure}

\section{Finite magnetic field}
We consider an undoped bilayer of graphene in the presence of an external
perpendicular magnetic field $B$.
 It is instructive to consider separately
the two gauges for the vector
potential ${\bf A}$ given below. The momentum operator ${\bf p}$, 
appearing in the term  $\mathcal{H}_0$ in Eq. (1), is shifted by $-e{\bf A}$.
\subsection{Landau gauge}
In this gauge  we have ${\mathbf A}=(0,Bx,0)$ with $[{\mathcal
H}_0,p_y]=0$. Consequently we can
write the
solutions for the spinor components in the form
$\psi_A(x,y)=\phi_A(x)e^{ik_y y}$, $\psi_B(x,y)=i\phi_B(x)e^{ik_y
y}$, $\psi_{A'}(x,y)=i\phi_{A'}(x)e^{ik_y y}$ and
$\psi_{B'}(x,y)=\phi_{B'}(x)e^{ik_y y}$. The $x$ dependence of the
spinor components is then described by
\begin{eqnarray}
&&\hspace*{-0.5cm}\hbar v_F\Bigl( \frac{d}{d x} + k_y+\beta x\Bigr)\phi_A =
(E-U_1)\phi_B,\\
&&\cr &&\hspace*{-0.5cm}\hbar v_F\Bigl(\frac{d}{d x} -
k_y-\beta x\Bigr)\phi_B = -(E-U_1)\phi_A + t\phi_{B'},\\
&&\cr &&\hspace*{-0.5cm} \hbar v_F\Bigl( \frac{d}{d x} +
k_y+\beta x\Bigr)\phi_{A'} = -(E-U_2)\phi_{B'}+t\phi_A,\\
&&\cr &&\hspace*{-0.5cm}\hbar v_F\Bigl(\frac{d}{d x} - k_y-\beta x\Bigr)\phi_{B'}
= (E-U_2)\phi_{A'},
\end{eqnarray}
where $\beta = eB/\hbar$. For a wide, uniform system these
equations can be decoupled to obtain, for $\phi_A$
\begin{equation}
\Bigl(\frac{d^2}{d \xi^2}+\alpha^2+\delta^2-\xi^2 \Bigr)^2\phi_{A}
=\{(\alpha^2-\delta^2) t'^2+[1-2\delta\alpha]^2\} \phi_{A}
\end{equation}
with $\xi = \ell_B k_y+x/\ell_B$; 
the characteristic length
scale of the system is
the magnetic length 
$\ell_B=
[\hbar/eB]^{1/2}$, $\alpha = \epsilon-u_0$ is the energy
shifted by the average potential between the layers
$u_0=(u_1+u_2)/2$, and $\delta=\Delta u/2$, $\Delta u = u_1-u_2$. The
energy, the potentials, and the interlayer coupling strength are
written in dimensionless units as $\epsilon = E\ell_B/\hbar v_F$,
$u_{1,2} = U_{1,2}\ell_B/\hbar v_F$ and $t' = t\ell_B/\hbar v_F$,
respectively.
Next, we we consider 
the equation
\begin{equation}
\Bigl(-
\partial^2/\partial \xi^2+\xi^2 \Bigr)\phi_{A}
=\gamma_{\pm}(\alpha)\, \phi_{A},
\end{equation}
where the potential-dependent eigenvalues are
\begin{equation}
\gamma_{\pm}(\alpha)=\alpha^2+\delta^2\pm
[(\alpha^2-\delta^2)
t'^2+(1-2\delta\alpha)^2]^{1/2}.
\end{equation}
With the substitution $\phi_A=f_A \exp(-\xi^2/2)$, and after some
straightforward algebra, the LL spectrum is given by the solutions
$\alpha_n$ of the fourth-order algebraic equation
\begin{equation}
[(\alpha+\delta)^2-2(n+1)][(\alpha-\delta)^2-
2n]-(\alpha^{2}-\delta^{2})
t'^2=0,
\end{equation}
where $n$ is an integer. The solutions of Eq. (11) can be obtained
in terms of Hermite polynomials.

Substituting   $\phi_A$ in Eqs. (6)-(9) we obtain
\begin{eqnarray}
&&\phi_A=d_n c_n\,H_n(\xi)\,e^{-\xi^2/2},\cr &&\cr
&&\phi_B=d_n\frac{\sqrt{2
n}}{(\alpha_n-\delta)}c_{n-1}\,H_{n-1}(\xi)\,e^{-\xi^2/2},\cr
&&\cr &&\phi_{A'}=-f_n d_n \frac{\sqrt{2
(n+1)}}{(\alpha_n+\delta)}c_{n+1}\,H_{n+1}(\xi)\,e^{-\xi^2/2},\cr
&&\cr &&\phi_{B'}=f_n d_n c_n\,H_n(\xi)\,e^{-\xi^2/2},
\end{eqnarray}
where $H_n(\xi)$ is the Hermite polynomial of order $n$ and
\begin{eqnarray}
&&\hspace*{-1.29cm}c_n=
1/
\left(n!2^n\sqrt{\pi}\ \right)^{1/2},\cr &&\cr
&&\hspace*{-1.29cm}f_n=\frac{[(\alpha_n-\delta)^2-2n]}{t'(\alpha_n-\delta)},\cr
&&\cr
&&\hspace*{-1.29cm}d_n=\Bigg[f_n^2\Bigl[1+\frac{2(n+1)}{(\alpha_n+\delta)^2}\Bigr]+1+\frac{2n}
{(\alpha_n-\delta)^2}\Biggr]^{-1/2}.
\end{eqnarray}
Notice that the eigenstates are determined by the quantum numbers
($k_y,n$) while the spectrum is independent of $k_y$.

\subsection{Symmetric gauge}
In problems with cylindrical symmetry it is advantageous to use
the symmetric gauge ${\bf A}=(0,B\rho/2,0)$.
 In this case the corresponding solutions for the four-component
spinors $\Psi$ are written as \cite{Vincenzo,Milton}
\begin{equation}
\Psi(\rho,\theta)=
\begin{pmatrix}
  \phi_A(\rho)e^{im\theta}  \\
  \ \\
  i\phi_B(\rho)e^{i(m-1)\theta}\\
\  \\
  \phi_{B'}(\rho)e^{im\theta}\\
\ \\
  i\phi_{A'}(\rho)e^{i(m+1)\theta}
\end{pmatrix}
,
\end{equation}
where $m$ is the angular momentum label. The radial dependence of
the spinor components is described, in dimensionless units, by
\begin{eqnarray}
&& \frac{1}{\sqrt{2}}\Bigl[\frac{d}{d \xi} + \frac{m}{\xi} +
\xi\Bigr]\phi_A = (\alpha-\delta)\phi_B,\cr &&\cr
&&\frac{1}{\sqrt{2}}\Bigl[\frac{d}{d \xi} - \frac{(m-1)}{\xi}-
\xi\Bigr]\phi_B = -(\alpha-\delta)\phi_A + t'\phi_{B'},\cr &&\cr
&& \frac{1}{\sqrt{2}}\Bigl[\frac{d}{d \xi} + \frac{(m+1)}{\xi}+
\xi\Bigr]\phi_{A'} = -(\alpha+\delta)\phi_{B'}+t'\phi_A,\cr &&\cr
&& \frac{1}{\sqrt{2}}\Bigl[\frac{d}{d \xi} - \frac{m}{\xi}- \xi
\Bigr]\phi_{B'} = (\alpha+\delta)\phi_{A'},
\end{eqnarray}
where now $\xi=\sqrt{2}\rho/(2\ell_B)$, and the other terms are
defined as before.

 Following the same procedure as in the previous case, one
can obtain an equation that is identical to that of an harmonic
oscillator
\begin{equation}
\Bigl[-\frac{1}{2}\Bigl(\frac{d^2 }{d \xi^2} +\frac{1}{\xi}\frac{d
}{d \xi} -\frac{m^2}{\xi^2}\Bigr)+m+
\frac{\xi^2}{2}\Bigr]\phi_A=2\gamma_{\pm}(\alpha)\phi_{A},
\end{equation}
where
\begin{equation}
\gamma_{\pm}(\alpha)=\Bigl(n'+1/2\Bigr),
\end{equation}
with $n'=n+(|m|+m)/2$, $n=0,1,2...$ and $\gamma_{\pm}(\alpha)$ is
given by Eq. (12). The LL spectrum is determined by
\begin{equation}
[(\alpha+\delta)^2-2(n'+1)][(\alpha-\delta)^2-2n']-(\alpha^2-\delta^2)
t'^2=0,
\end{equation}
which is identical to Eq. (13) except that here 
the Landau index
$n'$ is used instead of $n$.

It is evident, from Eq. (18) that the spinor components $\phi_A$
and $\phi_{B'}$ can be obtained in terms of generalized Laguerre
polynomials. For $m>0$ we obtain:
\begin{eqnarray}
&& \phi_A=C_{A n}^m\,
\xi^{|m|}e^{-\frac{\xi^2}{2}}L_n^{|m|}(\xi^2),\cr &&\cr &&\phi_B=
\frac{\sqrt{2n'}}{(\alpha_{n'}-\delta)}\,C_{A n}^{m-1}\,
\xi^{|m|-1}e^{-\frac{\xi^2}{2}}L_n^{|m|-1}(\xi^2),\cr &&\cr
&&\phi_{A'}=-\frac{\sqrt{2(n'+1)}}{(\alpha_{n'}+\delta)}\,C_{B'
n}^{m+1}\, \xi^{|m|+1}e^{-\frac{\xi^2}{2}}L_n^{|m|+1}(\xi^2),\cr
&&\cr && \phi_{B'}=C_{B' n}^m\,
\xi^{|m|}e^{-\frac{\xi^2}{2}}L_n^{|m|}(\xi^2).
\end{eqnarray}
For $m<0$ the result is
\begin{eqnarray}
&& \phi_A=C_{A n}^{m}\,
\xi^{|m|}e^{-\frac{\xi^2}{2}}L_n^{|m|}(\xi^2),\cr &&\cr &&\phi_B=
-\frac{\sqrt{2n'}}{(\alpha_{n'}-\delta)}\,C_{A n-1}^{m+1}\,
\xi^{|m|+1}e^{-\frac{\xi^2}{2}}L_{n-1}^{|m|+1}(\xi^2),\cr &&\cr
&&\phi_{A'}=\frac{\sqrt{2(n'+1)}}{(\alpha_{n'}+\delta)}\,C_{B'
n+1}^{m+1}\,
\xi^{|m|-1}e^{-\frac{\xi^2}{2}}L_{n+1}^{|m|-1}(\xi^2),\cr &&\cr &&
\phi_{B'}=C_{B' n}^{m}\,
\xi^{|m|}e^{-\frac{\xi^2}{2}}L_n^{|m|}(\xi^2),
\end{eqnarray}
where
\begin{equation}
C_{A n}^m=d_n/ 
\left[n!/(n+|m|)!\right]^{1/2},\ \ \ 
C_{B'n}^m = f_n
C_{A n}^m;
\end{equation}
 the other constants are as before.


\subsection{ Special cases}
 {\it i)} $t'=\delta=0$. This is the
case of a single graphene layer or two uncoupled, unbiased
layers. The solutions are then
\begin{equation}
\alpha_{n'} = \pm\sqrt{2(n'+1)},\qquad \alpha_{n'} =
\pm\sqrt{2n'},
\end{equation}
which are the well known expressions for the LL in a single-layer
of graphene.

{\it ii)} $\delta = 0$. That is the case of an unbiased bilayer.
The explicit solutions are
\begin{eqnarray}
&&\alpha_{n'} = \pm
\frac{t'}{\sqrt{2}}\Biggl[
1+\frac{2}{t'^2}(2n'+1) 
\cr &&\pm
\sqrt{\Bigl[1+\frac{2}{t'^2}(2n'+1)
\Bigr]^2-\frac{16}{t'^4}n'(n'+1)}\ \Biggr]^{1/2}.
\end{eqnarray}
This expression can be simplified by expanding the internal square
root. By taking the negative sign, one obtains
\begin{equation}
\alpha_{n'} = \pm
\frac{2}{t'}\sqrt{n'(n'+1)}/
\left[1+\frac{2}{t'^2}(2n'+1)\right]^{1/2}
\end{equation}
which is valid for $n'/t'^2<<1$. By assuming
$1/
[1+
(2/t'^2)(2n'+1)]^{1/2}\approx 1$, E2. (29)
is simplified to
\begin{equation}
\alpha_{n'} = \pm \frac{2}{t'}\sqrt{n'(n'+1)},
\end{equation}
which is the expression previously obtained in the context of the
two-band continuum model \cite{Falko}. By taking the positive sign
in front of the internal square root in Eq. (25), one finds the
higher-energy LL that arise from the upper band at $E = t$. For
$n'=0$ that gives
\begin{equation}
\alpha_0 = \pm \sqrt{t'^2+ 2}.
\end{equation}

\section{Numerical results for the energy spectrum} Figure 2 shows the first 6 positive energy LL as a
function of the magnetic field $B$ for an {\it unbiased} grahene
bilayer. The figure shows that as $B$ increases, Eq. (27)
consistently overestimates the values of the energy levels (dashed
lines) as compared to the exact results obtained from Eq. (25)
(solid) (which coincide with those from the approximate expression
Eq. (26)). The discrepancy increases with $B$
and, for the parameters usually observed experimentally ($t=0.4$
eV, $v_F \approx 1.0 \times 10^6$ m$/$s), can attain $5$ meV for
$B=15 T$ and $n=1$. This result was confirmed by a numerical
calculation based on a discretization of Eq. (17). Recent
experimental studies have measured the LL spectrum in bilayers of
graphene for fields up to $12$ T  \cite{Andrei}. At these fields,
the discrepancy between the four-band and two-band results are
$\approx 10 \%$ for the $n=1$ LL. This is below the energy
resolution of the current experimental techniques. However, for
stronger fields the difference should become large enough to
become detectable. In addition, the presence of the LL associated
with the higher energy band at $E \approx 0.4$ eV may be
detectable by the scanning tunneling technique discussed in ref.
23.
\begin{figure}
\includegraphics*[height=9.0cm, width=9.5cm]{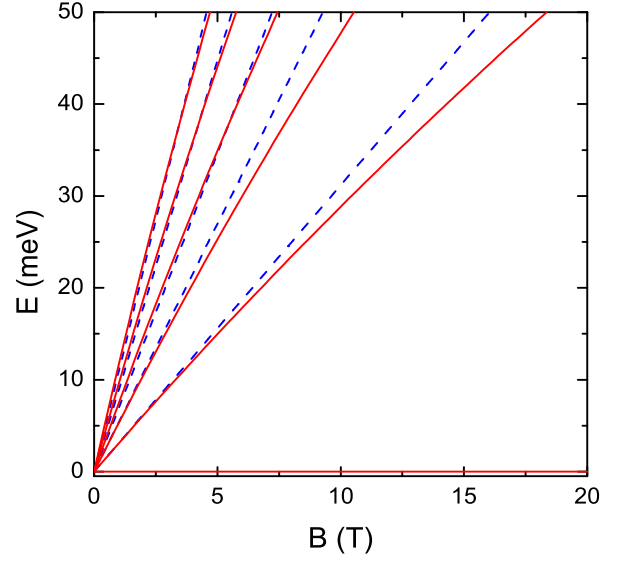}
\vspace*{-0.8cm} \caption{Landau levels $(n=0 - 5)$ in an {\it
unbiased}
 graphene bilayer
as a function of the magnetic field. The solid, red lines result
from a numerical solution of Eq. (25) and the dashed, blue ones
are the approximate results from Eq. (27).}
\end{figure}

Equation (20) shows that the charge-conjugation symmetry between
electrons and holes is preserved for the Landau states in a single
layer ($t=0$, $\delta=0$) as well as in the unbiased double layer
($\delta = 0$), since the equation is unchanged under the
transformation $\alpha \rightarrow -\alpha$. On the other hand,
that ceases to be the case for a {\it biased} bilayer of graphene.
Figure 3 shows the first 6 low-energy LL, for a biased bilayer
with $U_1=-U_2=50$ meV. The positive (negative) energy levels are
shown in Fig. 3a (3b), and the results show a non-monotonic
dependence on the magnetic field. To each index $n$ there
correspond four solutions, two negative and two positive, with the
low-energy branches being equal to $|\delta|$ for $B=0$, whereas
the high-energy solutions arise from the bands at $|\alpha| = t'$
(not shown). The only exception is the $n=0$ state, for which, in
the case of the $K$ valley, there is no positive solution at the
vicinity of $\delta$ (the opposite is true at the vicinity of the
$K'$ valley). The figure shows the LL corresponding to $n=0$
(black solid line), $n=1$ (red dashed lines), $2$ (blue dotted),
$3$ (green long-dashed), $4$ (orange dot-dashed) and $n=5$
(magenta dot-dot-dashed lines). One striking feature in the
spectrum is the appearance of crossings and energy minima in some
magnetic field regions shown in the insets. These features arise
as the potential difference (i.e. the gap parameter) between the
layers is increased, with the energy minima and the crossings
initially appearing at very small values of $B$. As $\Delta U$
increases, the minima in each branch ocurr at larger magnetic
fields and lower energies. These results can be explained by
taking into account the band structure of the biased graphene
bilayer. The introduction of a magnetic field in a 2D system
constrains the electronic states to closed cyclotron orbits, which
in turn causes an energy quantization in which the energy
eigenvalues correspond to momentum states that are integer
multiples of $h/\ell_B$. For a system with a parabolic band  this
gives rise to a linear dependence of the LL on the magnetic field.
For a single graphene layer, on the other hand, the linear
momentum dependence of the energy introduces a $\sqrt{B}$
dependence of the LL spectrum. In a biased graphene bilayer the
electron dispersion has a more complicated, "mexican hat" shape
(see Fig. 1) that becomes more pronounced as $\Delta U$ becomes
larger. For  $B=0$ this dispersion then allows the existence of
pairs of degenerate states with different values of momenta which,
for finite
$B$, can become degenerate LL with different Landau indices. 
By setting $n'=0$ in Eq. (20) we 
obtain, for $\alpha$ values close to $-\delta$,
\begin{equation}
\alpha_0 = -\delta\frac{(1-2\beta/t'^2)}{(1+2\beta/t'^2)}.
\end{equation}
The introduction of bias, together with the presence of a magnetic
field, breaks the inversion symmetry of the bilayer, which in turn
breakes the charge-conjugation symmetry. A similar effect is
obtained for the case of states in a position-dependent confining
potential \cite{Milton}. This electron-hole asymmetry is
particularly evident in the lowest LL levels, which are not
distributed symmetrically around $E=U_0$, i.e. $E_{n}\neq
-E_{-n}$, for $U_0 = 0$, except for $\Delta U \approx 0$.
\begin{figure}
\vspace*{-0.95cm} \hspace*{-0.65cm}
\includegraphics*[height=8.0cm, width=10.0cm]{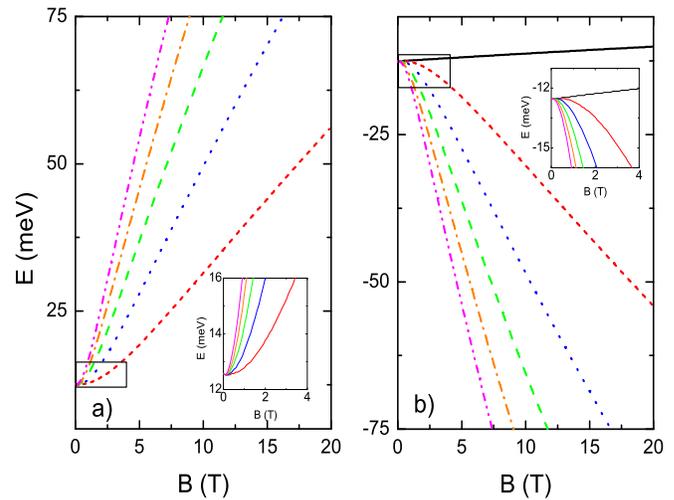}
\vspace*{-0.8cm} \caption{ a) Positive-energy Landau levels in a
{\it biased} graphene bilayer as a function of the magnetic field
$B$. (b) As in (a) for negative-energy levels. In both cases $U_1
= - U_2 = 50$ meV and $n=0$ (solid line), $n=1$ (dashed lines),
$2$ (dotted), $3$ (long dashed), $4$ (dash dotted) and $n=5$
(dot-dot-dashed lines). The rectangular areas are  enlarged in the insets.}
\end{figure}
\begin{figure}
\vspace*{-0.95cm} \hspace*{-1cm}
\includegraphics*[height=8.0cm, width=8.5cm]{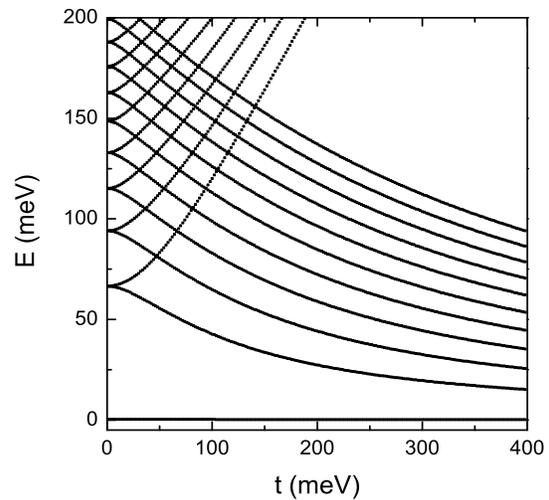}
\vspace*{-0.8cm} \caption{Landau levels in an {\it unbiased}
graphene bilayer,
for $B= 5$ T, as a function of the interlayer coupling
parameter $t$.}
\end{figure}

Previous studies of LLs in bilayer graphene that were based on a
continuum model have used a two-band approximation, which is
expected to be accurate for large coupling interlayer strengths
$t$. The four-band model used here allows one to investigate the
whole range of interlayer interaction strengths. Figure 4 shows
the first 10 LL for an unbiased graphene bilayer as a function of
the coupling parameter $t$. The $t=0$ limit corresponds to the
case of two uncoupled single-layers, and the spectrum is given by
Eq. (24). As the coupling increases, the two-fold degeneracy of
the levels is lifted, with the higher energy levels being shifted
towards $E=t$, and several crossings occur. The figure also shows
that the energy difference between the lowest LL decreases for
increasing $t$, with the $n=0$ LL remaining unaffected by the
coupling, in the absence of bias.

The dependence of the spectrum on the energy gap can be seen in
Fig. 5, which shows the LL corresponding to $n=0$ (black solid
line), $n=1$ (red dashed lines), $2$ (blue dotted), $3$ (green
long dashed), $4$ (orange dash dotted) and $n=5$ (magenta
dot-dot-dashed lines). One striking feature of this result is the
reduction of the energy difference between some of the LL as the
gap increases, with the eventual appearance of degeneracies. That
can be explained by considering the fact that as $\Delta U$
increases, the band structure of the bilayer evolves from an
approximately parabolic dispersion to the approximately
fourth-order polynomial dependence on the momentum. From Eq. (20)
one can easily show that degeneracies will occur between LL with
indices $n_1$ and $n_2$ for energies and fields that satisfy the
relation
\begin{equation}
n_1+n_2 = \alpha_{n_1}^2+\delta^2-1.
\end{equation}

\begin{figure}
\hspace*{-1cm}
\includegraphics*[height=8.5cm, width=8.5cm]{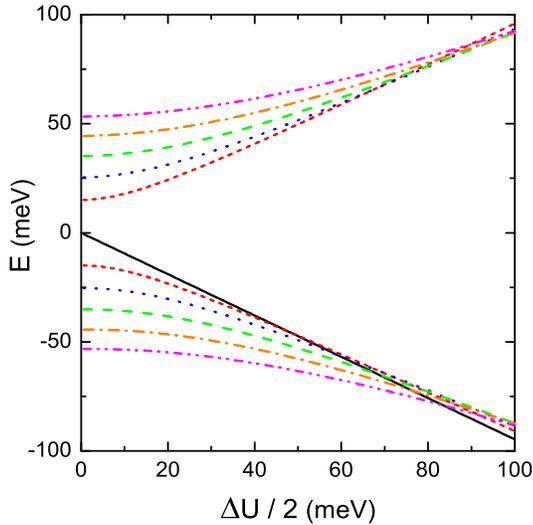}
\caption{Landau levels in a graphene bilayer  as a function of the
difference $\Delta U$ in potential between the layers  for $B = 5$
T: $n=0$ (black solid line), $n=1$ (red dashed lines), $2$ (blue
dotted lines), $3$ (green long-dashed lines), $4$ (orange
dash-dotted lines), and $n=5$ (magenta dot-dot-dashed lines).}
\end{figure}

\section{Cyclotron resonance oscillator strength}

The oscillator strength $|<\Psi^\dagger|\rho e^{i\phi}|\Psi>|^2$
for electric dipole transitions between the LL is given by
\begin{equation}
T=(T_{A}+T_{B}+T_{A'}+T_{B'})^2
\end{equation}
where
\begin{equation}
T_C=\int\phi_{C,l} \xi \phi_{C,l'} d\xi,
\end{equation}
with $C=A,B,A'$ and $B'$. These matrix elements were calculated in
the case of the Landau gauge, as
\begin{equation}
T_A=d_ld_{l'}\Bigl(\sqrt{l} \delta_{l,l'+1}+\sqrt{l+1}
\delta_{l,l'-1}\Bigr)/\sqrt{2},
\end{equation}
\begin{equation}
T_B=\frac{d_ld_{l'}2\sqrt{ll'}}{(\alpha_l-\delta)(\alpha_{l'}-\delta)}\Bigl(\sqrt{l-1}
\delta_{l,l'+1}+ \sqrt{l} \delta_{l,l'-1}\Bigr)/\sqrt{2},
\end{equation}
\begin{eqnarray}
\nonumber
T_{A'}&=&d_ld_{l'}f_lf_{l'}\frac{2\sqrt{(l'+1)(l+1)}}{(\alpha_l+\delta)(\alpha_{l'}+\delta)}\\*
&&\times\Bigl(\sqrt{l+1} \delta_{l,l'+1}+\sqrt{l+2}
\delta_{l,l'-1}\Bigr)/\sqrt{2},
\end{eqnarray}
\begin{equation}
T_{B'}=d_ld_{l'} f_lf_{l'}\Bigl(\sqrt{l}
\delta_{l,l'+1}+\sqrt{l+1} \delta_{l,l'-1}\Bigr)/\sqrt{2}.
\end{equation}
where $l$ and $l'$ are Landau indices of the different
eigenstates. The selection rule is $\delta l = |l|-|l'|=\pm 1$.
Here $(l,l')$ indicates a transition from a LL with index $l$ to
another with index $l'$; the negative sign is used to represent a
transition from a hole-like LL. The oscillator strengths for some
dipole-allowed transitions between LL are shown in Fig. 6, for a
biased bilayer (a) with $U_1=-U_2= 50$ meV and $B=10$ T, as well
as for an {\it unbiased} bilayer (b). The effect of the bias is
most clear in the shift in energy of the $(0,1)$ and $(-1,-2)$
transitions, caused by the gap in the spectrum.
On the other hand, one can also observe a decrease in
the energies associated with transitions between the higher LL.

Figure 7(a) shows the energy $\Delta E$ for some transitions
between LL in a biased graphene bilayer as a function of the
magnetic field, for $U_1=-U_2=50$ meV, namely $(0,1)$ (black
dashed lines), $(1,2)$ (blue dotted line), $(-1,-2)$ (black
solid), and $(-1,2)$ (red solid line). In contrast with the case
of the unbiased bilayer as well as the single layer, $\Delta E$ is
seen to be weakly dependent on the field for $B < 5$ T. For the
$(1,2)$ and $(-1,-2)$ transitions, the energy reaches negative
values, due to the existence of LL crossings (see Fig 3). In
addition, a difference in the energy of these transitions is found
(see inset). The results are obtained by subtracting the energies
associated with the initial and final LLs. However, recent
theoretical work has indicated that many-body corrections may
cause a substantial modification of the transition energies
\cite{Iyengar}. The right panel Fig. 7(b) shows the oscillator
strengths for the same transitions as in the left panel, as a
function of the magnetic field. These results show that, for small
fields, e.g. for  $B < 10$ T, the oscillator strength associated
with the $(-1,2)$ transition is slightly larger than that of the
$(0,1)$ transition, whereas for larger fields the latter becomes
stronger. The curve for the $(-1,2)$ transition intercepts the
result for the $(0,1)$ one at $B \approx 11.6$ T, which
corresponds to $E_1=-U_1$ (see Fig. 3(b)). As in Fig. 7(a), the
results show a significant asymmetry between the electron (1,2)
and hole (-1,-2) intraband transitions.

This difference between the electron and hole intraband
transitions becomes larger as the gap parameter ($\Delta U$) is
increased, as shown in Fig. 8, which displays the energies for the
$(1,2)$ (solid blue line) and $(-1,-2)$ (dashed red line)
transitions as a function of $\Delta U$ for $B = 20$ T, and
$(1,2)$ (solid green line) and $(-1,-2)$ (dotted black line) for
$B = 10$ T. At $\Delta U = 200$ meV the energy difference reaches
$\approx 3$ meV for $B = 20$ T, and $\approx 0.8$ meV for $B = 10$
T. For the $B =10$ T case, the energy difference reaches zero and
eventually becomes negative, due to the crossing of the LL. The
asymmetry between the electron and hole intraband transitions is
also particularly evident in their oscillator strength values, as
shown in Fig. 9, which presents the oscillator strengths for the
transitions discussed in Fig. 7, as a function of the gap
parameter, for $B=5$ T. The results show that the presence of a
gap causes a significant reduction of the oscillator strength for
the $(0,1)$ and $(1,2)$ transitions, whereas the result for
$(-1,2)$ shows a slight increase at small fields, reaching a
maximum at $\Delta U \approx 25$ meV with a subsequent decrease.
On the other hand, the curve for the $(-1,-2)$ transition has a
striking increase in comparison with the result for $\Delta U =
0$.

Figure 10(a) shows the energies for the $(1,2)$ and $(-1,-2)$
transitions (i.e. the cyclotron resonance transitions) as a
function of field, for several values of the gap parameter:
$\Delta U = 0$ (black solid line), $\Delta U = 25$ meV (red dashed
lines), $50$ meV (blue dotted lines), $100$ meV (green dot-dashed
lines) and $\Delta U = 200$ meV (magenta dot-dot-dashed lines).
For $\Delta U = 0$, the $\Delta E$ for the two transitions
coincides, whereas in the other cases, the energy associated with
the hole intraband transition is always higher than the
corresponding electron transition. Note that the $\Delta U = 0$
result only exhibit linear $B$-dependence, as predicted by the
simple expression Eq. (27), for small $B$-fields.
With increasing $\Delta U$ the cyclotron energy decreases and for
small $B$-values $\Delta E$ becomes even negative, indicating a
reversion of the order of the LL. These qualitative changes in
$\Delta E$ should be observable in a cyclotron resonance
experiment.

The oscillator strengths for the transitions shown in the left
panel are displayed in Fig. 10(b). It is seen that for the gapless
system the oscillator strengths are independent of the magnetic
field. On the other hand, as the gap increases, the results show a
strong asymmetry that increases as $\Delta U$ becomes larger. In
all cases, as $B$ increases, the results approach the oscillator
strengths for the $\Delta U = 0$ case. A similar behavior is also
found for the other $(l,l+1)$ and $(-l,-l-1)$ transitions, albeit
with smaller discrepancy in the energies.

The asymmetry between the electron and hole intraband transitions
can be explained by taking into account the functional dependence
of the amplitude factors of the spinor components on the energy.
These terms, in the presence of a gap, introduce an asymmetry
between the envelope functions of electrons and holes, thus
modifying the transition probabilities. This effect is in turn a
consequence of the breaking of the inversion symmetry of the
biased bilayer.

\begin{figure}
\vspace*{-0.95cm} \hspace*{-0.5cm}
\includegraphics*[height=7.5cm, width=9.5cm]{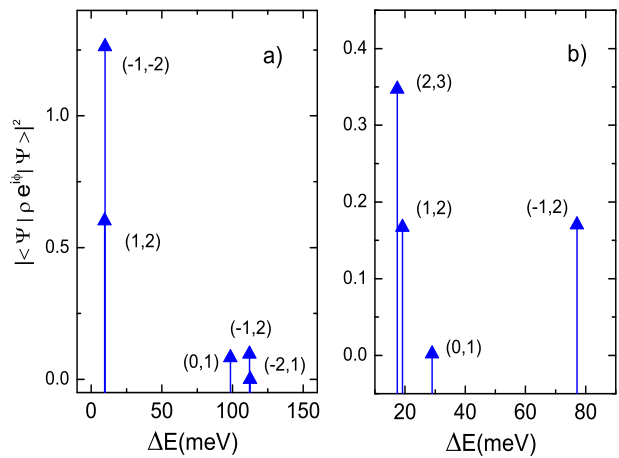}
\caption{(a) Oscillator strengths for electric dipole transitions in
 a {\it biased} bilayer of graphene ($\Delta U=100$ meV) for  $B=10$
T.
(b) As in (a) for an {\it unbiased} ( $\Delta U=0$ meV) bilayer. }
\end{figure}
\begin{figure}
\vspace*{-0.85cm} \hspace*{-0.5cm}
\includegraphics*[height=7.5cm, width=9.5cm]{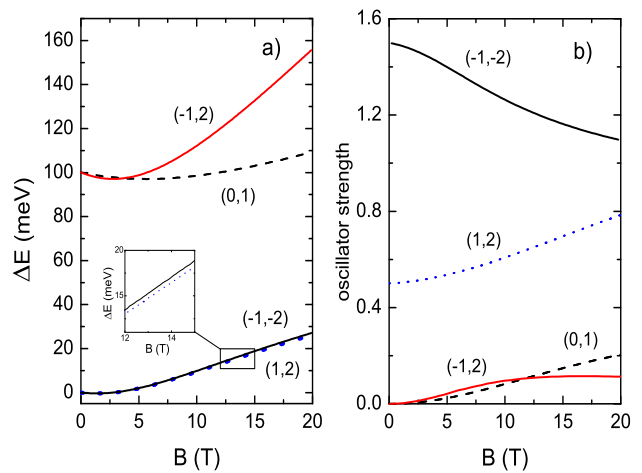}
\vspace*{-0.8cm} \caption{(a) Transition energies  in a {\it
biased} ($\Delta U=100$ meV) graphene bilayer, as a function of
the magnetic field $B$, for the dipole-allowed transitions $(0,1)$
(black dashed line), $(1,2)$ (blue dotted line), $(-1,-2)$ (black
solid line), and $(-1,2)$ (red solid line). (b) Oscillator
strengths vs field $B$ for the transitions
described in (a).} 
\end{figure}
\begin{figure}
\vspace*{-0.9cm}
\includegraphics*[height=8.5cm, width=9cm]{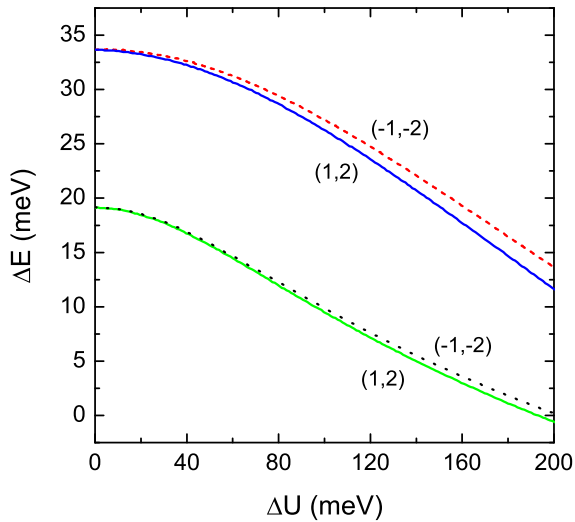}
\vspace*{-0.8cm} \caption{Transition energies as a function of
$\Delta U$ for the (1,2) (blue solid line) and (-1,-2) (red dashed
line) transitions for $B = 20$ T, and (1,2) (green solid line) and
(-1,-2) (black dotted line) for $B = 10$ T.}
\end{figure}
\begin{figure}
\vspace*{-0.9cm}
\hspace*{-1cm}
\includegraphics*[height=8.0cm, width=8.5cm]{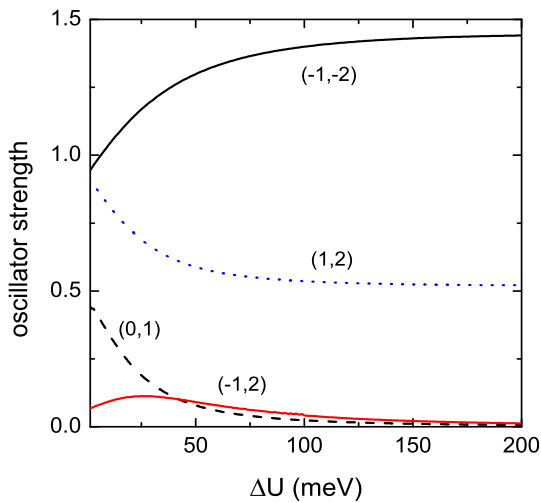}
\vspace*{-0.8cm} \caption{Oscillator strengths for dipole-allowed
transitions in a graphene bilayer as a function of the interlayer
potential difference $\Delta U$ for the $(0,1)$ (black dashed
line), $(1,2)$ (blue dotted line), $(-1,-2)$ (black solid) and
$(-1,2)$ (red solid line) transitions, at $B=5$ T.}
\end{figure}
\begin{figure}
\vspace*{-0.9cm} \hspace*{-0.8cm}
\includegraphics*[height=8.2cm, width=10.0cm]{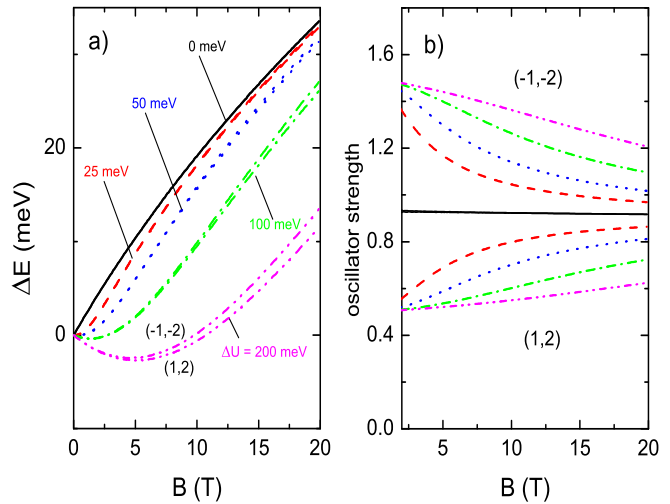}
\vspace*{-0.8cm} \caption{(a) Energies of the $(1,2)$ and
$(-1,-2)$ (upper curves of each doublet) transitions as a function
of magnetic field for $\Delta U = 0$ (black solid line), $\Delta U
= 25$ meV (red dashed lines), $50$ meV (blue dotted lines), $100$
meV (green dot-dashed lines) and $\Delta U = 200$ meV (magenta
dot-dot-dashed lines). (b) The oscillator strengths for the
transitions shown in (a) as a function of magnetic field. The
results below $0.92$ correspond to the $(1,2)$ transitions.}
\end{figure}

\section{Summary and Conclusions}

Using a four-band model we  obtained analytical expressions
for the Landau levels and the eigenfunctions  in a
{\it biased} graphene bilayer in the  presence of an external
perpendicular magnetic field $B$.
 In doing so we also extended previous results for the discrete spectrum of an {\it unbiased} graphene layer
 based on a two-band model \cite{Falko}. Further, we obtained the
selection rules for electric dipole transitions and evaluated the
associated oscillator strengths.

Relative to an {\it unbiased} bilayer the introduction of a bias
modifies  the spectrum considerably by reducing the  energy for
transitions between two electron states while at the same time
increasing the energy for transitions between an electron and a
hole state, see Figs. 5 and 6. Also noteworthy is the modification
of the electronic band structure of the bilayer caused by the
bias, which leads to the breaking of electron-hole symmetry, the
appearance of degeneracies in the LL spectrum, and qualitative
different dependency of the transition energy $\Delta E$ on the
magnetic field, which in turn can strongly modify the dipole
transition probabilities. These effects can be probed by cyclotron
resonance measurements \cite{Deacon}, scanning tunneling
spectroscopy \cite{Andrei} as well as by far-infrared spectroscopy
\cite{Potemski}. Future work may take into account the effects of
spin as well as electron-electron interactions, which may become
particularly important at high magnetic fields
\cite{Stauber,Jiang}.

\section{Acknowledgements}
 This work was supported by the Brazilian
Council for Research (CNPq), BOF-UA, the Flemish Science
Foundation (FWO-Vl), the Belgian Science Policy (IAP), and the
Canadian NSERC Grant No. OGP0121756.

\end{document}